\documentclass[sigconf]{acmart}
\usepackage{graphicx}

\AtBeginDocument{%
  \providecommand\BibTeX{{%
    \normalfont B\kern-0.5em{\scshape i\kern-0.25em b}\kern-0.8em\TeX}}}

\setcopyright{acmlicensed}
\copyrightyear{2024}
\acmYear{2024}
\acmDOI{XXXXXXX.XXXXXXX}

\acmConference[SIGGRAPH '24 Posters]{Special Interest Group on Computer Graphics and Interactive Techniques Conference Posters (SIGGRAPH ’24 Posters)}{July 28--August 1}{Denver, CO}

\setcopyright{none}
\acmConference[preprint for SIGGRAPH'24 Posters]{preprint for Special Interest Group on Computer Graphics and Interactive Techniques Conference Posters (SIGGRAPH ’24 Posters)}{July 28--August 1}{Denver, CO}

\acmBooktitle{SIGGRAPH ’24 Posters} 

\usepackage{xspace}

\newcommand{\commentout}[1]{}

\begin{document}

\title{Measurement of the Imperceptible Threshold for Color Vibration Pairs Selected by using MacAdam Ellipse}

\author{Shingo Hattori}
\affiliation{%
  \institution{Cluster Metaverse Lab}
  \streetaddress{8-9-5 Nishigotanda, Shinagawa}
  \city{Tokyo}
  \country{Japan}}
\affiliation{%
  \institution{University of Tsukuba}
  \streetaddress{1-2 Kasuga, Tsukuba}
  \city{Ibaraki}
  \country{Japan}}
\email{s.hattori@cluster.mu}

\author{Yuichi Hiroi}
\affiliation{%
  \institution{Cluster Metaverse Lab}
  \streetaddress{8-9-5 Nishigotanda, Shinagawa}
  \city{Tokyo}
  \country{Japan}}
\email{y.hiroi@cluster.mu}

\author{Takefumi Hiraki}
\affiliation{%
  \institution{Cluster Metaverse Lab}
  \streetaddress{8-9-5 Nishigotanda, Shinagawa}
  \city{Tokyo}
  \country{Japan}}
\email{t.hiraki@cluster.mu}

\renewcommand{\shortauthors}{Hattori, et al.}

\begin{abstract}
We propose an efficient method for searching for color vibration pairs that are imperceptible to the human eye based on the MacAdam ellipse, an experimentally determined color-difference range that is indistinguishable to the human eye.
We created color pairs by selecting eight colors within the sRGB color space specified by the ellipse, and conducted experiments to confirm the threshold of the amplitude of color vibration amplitude at which flicker becomes imperceptible to the human eye.
The experimental results indicate a general guideline for acceptable amplitudes for pair selection.
\end{abstract}


\begin{teaserfigure}
  \includegraphics[width=\textwidth]{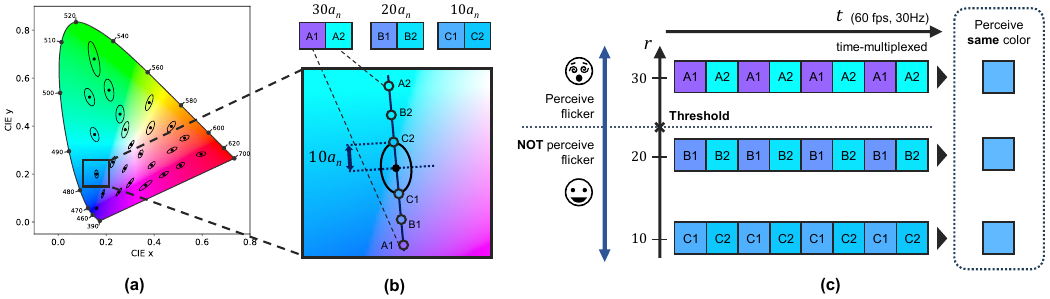}
  \caption{
  (a) MacAdam Ellipse~\cite{MacAdam1942-sv}, the region that encompasses all colors that cannot be distinguished by the average human eye from the color in the center of the ellipse on the $xy$ chromaticity diagram. For illustrative purposes, the ellipses are depicted as 10 times their actual size.
  (b) The two points of each McAdam ellipse, whose major diameter $a_n$ is multiplied by $r$, are selected for color vibration. 
  (c) Users view the color vibration on a color-calibrated display. The two colors selected by (b) are displayed alternately. In every pair, the human eye perceives the intermediate color resulting from the fusion of the two colors.
}
  \label{fig:teaser}
\end{teaserfigure}


\maketitle

\section{Introduction}
Imperceptible color vibration is a perceptual phenomenon in which the human visual system interprets the rapid alternating presentation of two colors with the same luminance but different chromaticity as a single fused intermediate color~\cite{Jiang2007-sh}.
This effect occurs when the alternation frequency exceeds the critical color fusion frequency (CCFF), which is typically around 25~Hz, leading to the perception of a blended color without discernible flicker~\cite{Truss1957-um}.

Imperceptible color vibration can be used to embed information such as a 2D barcode in LCD images that is not perceived by the human eye but detectable by cameras or photosensors~\cite{Abe2020-md}.
An increase in the amplitude of the color vibration facilitates the detection of the vibration by the camera, yet simultaneously renders the flicker more perceptible to humans.
The conventional approach to selecting color-vibration pairs is based solely on distances in the color space~\cite{Abe2020-md, Hattori2022-em}.
However, the human eye exhibits inequality in color perception; for instance, even if two colors on the chromaticity diagram are separated by the same distance, two colors closer to blue can be distinguished, but two colors closer to green cannot. 
The method of efficiently searching for and generating imperceptible color vibrations taking into account the perceptual characteristics of the human eye is not explored.

This paper proposes a method for selecting color vibration pairs with respect to the inequality of color perception based on the MacAdam ellipse~\cite{MacAdam1942-sv} (Fig.~\ref{fig:teaser}).
The MacAdam ellipse represents the experimentally determined range of color differences that cannot be distinguished by the human eye with respect to a particular color.
We conducted the experiment to create color pairs based on this ellipse and to confirm the threshold of the amplitude of the color vibration amplitude at which flicker cannot be detected by the human eye.

\global\long\def\Matrix#1{\mathbf{\MakeUppercase{#1}}}
\global\long\def\Vector#1{\mathbf{\MakeLowercase{#1}}}
\global\long\def\Transpose{\mathrm{T}}

\section{Methods}
The MacAdam ellipse $\mathcal{E}_n=\{\Vector{c}_n, \theta_n, a_n, b_n\}~(n=1\cdots 25)$ is defined at 25 points on the $xy$ chromaticity diagram by its center $\Vector{c}_n=[c_{nx},c_{ny}]$, rotation angle $\theta_n$, and the lengths $a_n$ and $b_n$ of long and short diameter. 
We choose color pairs $\{\Vector{p}^{+}_n(r), \Vector{p}^{-}_n(r)\}$ multiplied by ratio $r$ along the long diameter $a_n$ of these ellipses as color vibration pairs, denoted as $\Vector{p}^{\pm}_n(r)=[c_{nx}\pm r\cdot a_n \sin\theta_n,~  c_{ny}\pm r\cdot a_n \cos\theta_n]$.
This allows for the selection of color vibration pairs while taking into account the inequality of human color perception.

$xy$ chromaticity diagram is calculated by normalizing the luminance $Y$ of the CIEXYZ color space.
Therefore, when displaying colors based on the selected $\Vector{p}_n^{\pm}$ in practice, it is necessary to complement the luminance.
As $Y$ approaches 0, $\Vector{p}^{\pm}_n$ approaches black, and colors become nearly invisible. In contrast, as $Y$ approaches 1, the brightness of $\Vector{p}^{\pm}_n$ exceeds the sRGB range.
Therefore, we set $Y=0.4$ and convert $xy$ value to XYZ with \texttt{xyY\_to\_XYZ(xyY)} from the colour-science library in Python~\footnote{Colour 0.4.4 by Colour Developers, https://zenodo.org/records/10396329}.
The selected color pairs in CIEXYZ are then converted to the sRGB color system to display the pairs.
To convert the colors, we use CIE 1931 2$^\circ$ as an observer function under the D65 illuminant.

\section{Experiment and Results}
\begin{figure}[h]
  \centering
  \includegraphics[width=0.46\textwidth]{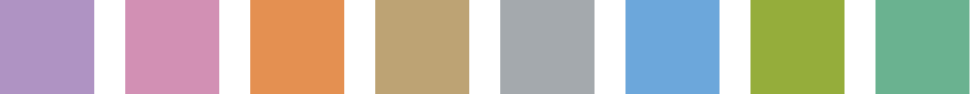}
  \caption{Images of 8 colors defined by the MacAdam ellipses used to generate color vibrations in the experiment.}
  \label{fig:colors}
  \vspace{-3mm}
\end{figure}

We conducted a user experiment to confirm the threshold of $r$ at which humans perceive flicker.
In this experiment, 8 points within the sRGB color space were selected from 25 points within the $xy$ color space defined by the MacAdam ellipse to generate color vibrations.
Figure~\ref{fig:colors} shows the colors of the 8 selected points.
We generate color pairs with $r$ values from 1 to around 40, split into 8 varying intervals across colors, and randomly presented these pairs on a screen in front of the participant.
Since the color space in sRGB is smaller than the $xy$ color space, some of $\Vector{p}_n^{\pm}$ cannot be reproduced in the sRGB color space.
By omitting these pairs, we used 46 pairs in this experiment.
To detect random responses, 46 single colors without color vibration were also displayed. As a result, each participant saw a total of 92 color patterns.


Ten participants (8 males, 2 females) are asked to respond if they can perceive flicker from the displayed colors.
The participants were seated in front of a sRGB color-calibrated LCD display (ColorEdge CG2420-Z, EIZO Inc.) at a distance of 60 cm so that their eye level was in the center of the monitor.
Each color pair was displayed as a 15 cm square at 60 cm from the monitor.
After each of the five pairs, the participants took a break to look at the black screen.
The protocol was approved by Cluster, Inc. Research Ethics Committee, and informed consent was obtained from all participants.

\begin{figure}[h]
  \centering
  \includegraphics[width=0.46\textwidth]{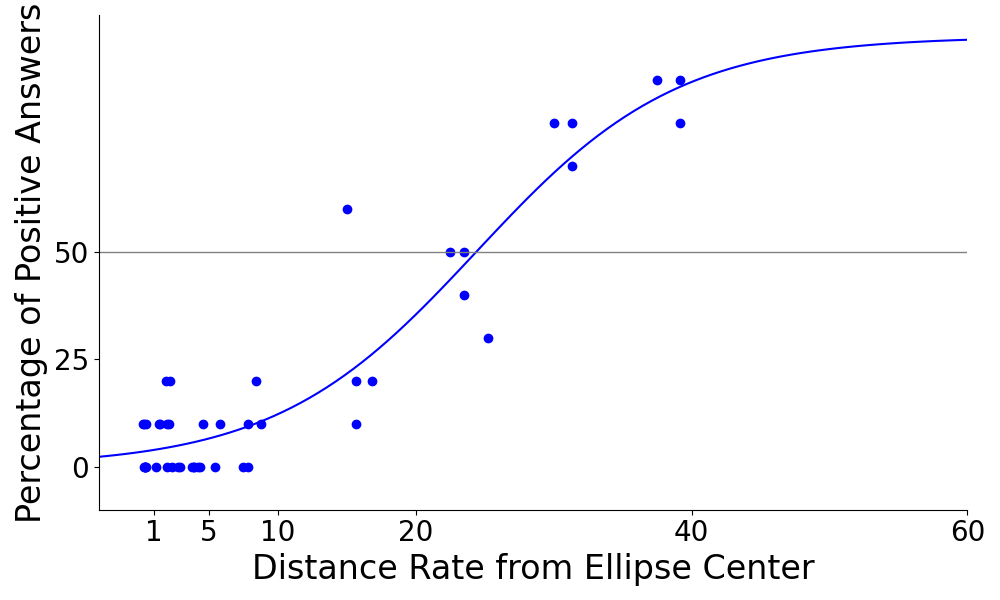}
  \caption{Percentage of responses perceived flicker relative to $r$. The blue dots indicate the mean, and the blue line shows the perceptual curve fitted by the sigmoid function.}
  \label{fig:results}
  \vspace{-3mm}
\end{figure}

Figure~\ref{fig:results} shows the average percentage of responses perceived as flicker compared to $r$.
From the result, the $r$ value of perceiving color vibration 50~\% of the time was $r=24.4$.
The results show that perceptual tolerance is approximately $\times~24$ expanded when the two colors alternate in time, in contrast to the MacAdam ellipses, which examined discrimination thresholds when the two colors are juxtaposed spatially.

\section{Conclusion}
This paper presented a method for generating color vibration pairs based on perceptual metrics and a general policy on acceptable amplitudes for pair selection.
This method of selecting color pairs can be extended to any color by interpolating MacAdam ellipses at each point. Future studies include the threshold of $r$ relative to the hue value and the adaptation of individual perceptual differences.


\bibliographystyle{ACM-Reference-Format}
\bibliography{sample-base}

\appendix

\end{document}